\documentclass[prl,aps,tightline,twocolumn,nofootinbib,showpacs,preprintnumbers]{revtex4-1}
\usepackage{amsmath} 
\usepackage{amsfonts} 
\usepackage{amssymb}
\usepackage{graphicx}
\usepackage{color} 
\newcommand {\be}{\begin{eqnarray}}
\newcommand{\ee}{\end{eqnarray}}
\def\n{\nonumber \\}
\begin{document}

\preprint{KEK-TH-1758}
\preprint{KEK-Cosmo-152}

\title{Small-field Coleman-Weinberg inflation driven by a fermion condensate} 
\author{Satoshi Iso} 
\author{Kazunori Kohri}
\author{Kengo Shimada}

\affiliation{
Institute of Particle and Nuclear Studies, High Energy Accelerator
Research Organization(KEK),  
Oho 1-1, Tsukuba, Ibaraki 305-0801, Japan \\
Graduate University for Advanced Studies (SOKENDAI), \\
Tsukuba 305-0801, Japan
}

\begin{abstract} 
We revisit the small-field Coleman-Weinberg inflation, which
 has the following two problems:
First, the smallness of the slow roll parameter $\epsilon$ requires
the inflation scale to be very low.  Second, the spectral index $n_s
\approx1+2 \eta$ tends to become smaller compared to the observed
value.  In this paper, we consider two possible effects on the dynamics of inflation:
radiatively generated nonminimal coupling to gravity $\xi
\phi^2 {\cal R}$ and condensation of fermions coupled to the
inflaton as $\phi \bar\psi \psi$.
We show that the fermion condensate can solve the above problems.
\end{abstract}

\maketitle


{\it Introduction}.--- The discovery of the standard model (SM) Higgs
boson as well as strong constraints on the supersymmetric parameters
forces us to reconsider the basic principles of particle physics.  In
particular the naturalness problem \cite{naturalness} of the
electroweak (EW) scale  has received renewed  interest.  
The classical conformality principle was advocated by
B. Bardeen as an alternative solution to the naturalness problem, and
various extensions of the SM based on the Coleman-Weinberg (CW)
mechanism\cite{CW} are proposed.  Since the CW mechanism does not work
within the SM due to the large top Yukawa coupling, we anyway need an
additional scalar sector in which the symmetry is radiatively broken
via the CW mechanism, which triggers the EW symmetry breaking.
The theoretical consistency with the
naturalness of the EW scale indicates that the breaking scale $M$ in
the additional sector must be much lower than the Planck scale $M_{Pl}.$

Two types of inflations are possible if the inflaton field $\phi$ has
the CW potential: the large-field inflation (LFI) and the small-field
inflation (SFI).
The large field type of the CW inflation is widely
studied as a special case of the chaotic inflation models.
On the other hand, the small-field CW inflation was studied in the early 1980s in the nonsupersymmetric GUT
models \cite{Linde:1981mu,Albrecht:1982wi,Shafi:1983bd}.  Suppose that
the inflaton field is trapped at the origin due to thermal corrections
to the effective potential generated in the reheating of the LFI.
When the fluctuations of the field are dominated by the vacuum
energy at $\phi=0$, the second inflation occurs,
and the radiation generated so far is rapidly diluted.  Then the inflaton field $\phi$
starts to roll down  to the true minimum at $\phi=M$.
Since the slow roll parameters in the SFI satisfy $\epsilon \ll
|\eta|$, the amplitude of the scalar perturbations $\Delta_R^2 =
V/24\pi^2 M_{Pl}^4 \epsilon$ becomes very large unless the vacuum
energy $V$ is sufficiently small.  The problem can be solved by
setting the scale of the inflation much lower than the GUT scale, but
then the e-folding number of the inflation is lowered,
and accordingly the spectral index $n_s$ of the scalar perturbation becomes smaller than
$0.94$, which deviates from the current observational bound, $n_s=
0.942-0.976$ \cite{Ade:2013zuv}.
The purpose of the present paper is to give a way to reconcile the predicted $n_s$ in the small-field CW inflation with the
observation.
In particular, we study  two effects on the dynamics of SFI:  nonminimal
coupling to gravity and  condensation of fermions coupled to the inflaton field $\phi$.
\\
\par
{\it Small-field CW inflation}.--- 
We first give a brief summery of the small-field CW inflation and its inherent problem pointed out in Refs.\cite{EJChun,Takahashi:2013cxa}. 
The CW potential
is given by
\be
V(\phi) = \frac{A}{4} \phi^4 \left( \ln\frac{\phi^2}{M^2}-\frac{1}{2} \right) + V_0, \ \ V_0=\frac{A M^4}{8} \  .
\label{CWpotential}
\ee
The choice of  the constant $-1/2$ in the bracket corresponds to 
taking the renormalization condition $V'(M) =0$ at the scale $\phi = M$.\footnote{
The mass-dependent renormalization scheme can be adopted 
which improves the calculation of the effective action by taking
into account  the threshold effects and hence the nonlogarithmic corrections to the potential \cite{Berera}.
In the present paper, 
since  particles are massless near the origin of the potential,
 the threshold corrections are expected not to become so large,
and we adopt the mass-independent renormalization scheme. 
But, as the slow roll parameters
and other cosmological quantities are very sensitive to  details of the potential, it may be
important to take these effects more carefully.  
}
$V_0$ is determined so that $V(M)=0$.
In this paper, we assume  $M \ll M_{Pl}=2.4 \times 10^{18}$ GeV.  
The
quartic coupling and its $\beta$ function at the scale $M$ are given
by $6 \lambda=V^{(4)}(M)=22A $ and $\beta_\phi=2A$, respectively.  Taking derivatives with respect to $\phi$, we have 
\be V' = A \phi^3 \ln
\frac{\phi^2}{M^2} , \ V'' = A \phi^2 \left( 2 + 3 \ln
  \frac{\phi^2}{M^2} \right) .
\ee
Hence, the inflaton mass is given by $m_\phi^2=V''(M)=\beta_\phi M^2$.\footnote{
The effective potential is parametrized by two quantities $A$ and 
$M$. Instead we can say that the potential is determined by two physical quantities,
the renormalized coupling at the minimum and the mass. 
In the small-field inflation, we show that the slow roll parameters depend on both of
these two quantities. In the $\phi^4$ large-field inflation, however, 
the renormalization scale dependence cancels out in the slow roll paramters \cite{Enqvist}.
We thank the referee for pointing out the difference.
}
The slow roll parameters are calculated to be
\begin{eqnarray}
\epsilon &=& \frac{M_{\rm Pl}^2}{2} \left(\frac{V'}{V} \right)^2
 \approx 32 \left( \frac{M_{\rm Pl}}{M} \right)^2 
\left( \frac{\phi}{M} \right)^6 \left( \ln \frac{\phi^2}{M^2} \right)^2 \ , 
\\
 \eta &=& M_{\rm Pl}^2 \left( \frac{V''}{V} \right) \approx 
24 \left( \frac{M_{\rm Pl}}{M} \right)^2
\left( \frac{\phi}{M} \right)^2  \ln \frac{\phi^2}{M^2}  \  .
\label{eta}
\end{eqnarray}
Here we used $V \approx V_0$ in the region $\phi \ll M.$
The slow roll conditions $\epsilon, |\eta| <1$  require that
the field value $\phi$ during inflation
must be much smaller than $M$.\footnote{
Generally speaking, the running of the coupling and higher-order corrections need to be
incorporated in studying the physical properties of inflation \cite{Enqvist}.
In our model, although the smallness of the field value $\phi \ll M$ gives $\ln (M/\phi) \sim 10^{1-2}$,
higher-order corrections are still negligible because various couplings are supposed to be very small in order to reproduce the observed amplitude of the scalar perturbation (\ref{scalaramplitude}).
Therefore the 1-loop CW potential (\ref{CWpotential})  is sufficient to calculate the physical quantities near the
origin.
}

Then the relation $\epsilon \ll |\eta|$ is satisfied.
Inflation stops at  $|\eta|=1$, where the slow roll condition is violated.
Equation (\ref{eta}) can be approximately solved as
\be
\phi^2/M^2 &\approx& (|\eta|/24 \ln(24 M_{Pl}^2/|\eta|M^2)) (M/M_{Pl})^2 \n
 &\approx& 10 ^{-3} |\eta| (M/M_{Pl})^2  \ll 1.
 \label{fieldvalue}
\ee
In the last equality, we set $M=10^{10} {\rm GeV}$ and $|\eta|=0.02$, but
the coefficient $10^{-3}$ is insensitive to these values.
 The slow roll parameter $\epsilon$ is given by
\be
\epsilon =  \frac{|\eta|^3}{432 \ \ln(24 M_{Pl}^2/|\eta|M^2) }  \left(\frac{M}{M_{Pl}} \right)^4 \ll 1 \  .
\ee
In order to make the amplitude of the scalar perturbation 
\be
\Delta_R^2 \approx \frac{V_0}{24 \pi^2 M_{\rm Pl}^4 \epsilon} 
= \frac{9 A \ln(24 M_{Pl}^2/|\eta|M^2)}{4 \pi^2 |\eta|^3} 
\label{scalaramplitude}
\ee
consistent with the Planck data \cite{Ade:2013zuv}, $\Delta_R^2 =
2.215 \times 10^{-9}$ at the pivot scale $k_0=k_{\rm CMB}=0.05$
Mpc$^{-1}$, the coefficient $A$ must be extremely small
$A \sim 10^{-15}$.
Hence, the potential height is given by
$V_0^{1/4} \sim 10^{-4} M$. Hereafter, the subscript ``CMB'' means the
value evaluated at the pivot scale $k=k_{\rm CMB}$.

The e-folding number $N$ is given by
\be
N &=& \frac{1}{M_{\rm Pl}^2} \int_{\phi_{end}}^{\phi} \frac{V}{V'} d \phi
\approx \frac{3 }{2  }
\left( \frac{1}{|\eta|} -  \frac{1}{|\eta_{end}|} \right) \  .
\ee
By setting $|\eta_{end}|=1$, we have   $\eta=-1/(2N/3+1).$
Since $\epsilon \ll |\eta|$, the spectral index  of the scalar perturbation is given by
$n_s=1+2 \eta$.
Hence, $n_s\sim 0.96$ \cite{Ade:2013zuv} requires a large e-folding number 
$N=3/(1-n_s)-3/2=73.5$ in the small-field CW inflation.

On the other hand, 
the e-folding number at the pivot scale of CMB measurement is given by
\begin{eqnarray}
   \label{eq:Ncond}
   N_{\rm CMB} &=& 61 + \frac23 \ln  \left(\frac{V_0^{1/4}}{10^{16} \mbox{GeV}} \right)
   + \frac13 \ln \left( \frac{T_R}{10^{16} \mbox{GeV}} \right),
   \label{CMBefolding}
\end{eqnarray}
where we assume that there was an epoch of the inflaton field's
oscillation induced by its mass term after the inflation and before
the reheating. After the reheating, we also assume that the radiation-dominated epoch continues until the matter-radiation equality epoch.
The smallness of the vacuum energy $V_0^{1/4} \sim 10^{-4} M \ll
M_{Pl}$ suggests a small e-folding number, which is inconsistent with
the above large e-folding number $N=73.5$.
The authors \cite{EJChun} considered a brane world scenario to reconcile the prediction with the
measurement \footnote{See also Ref. \cite{Nakayama:2012dw} for a
  solution by using supersymmetry.}.
In this letter, we study the
following two effects on the dynamics of inflation: a negative
nonminimal coupling to gravity and condensations of fermions.  The
first gives a negative quadratic term $-6|\xi| \phi^2$ in $V(\phi)$
while the second induces a linear term $-C \phi$.\footnote{It is
  known  to increase $n_s$ in the context of
the discrete $R$-symmetry
  models based on supersymmetry.~\cite{Takahashi:2013cxa}
}
\\
\par
{\it CW inflation with nonminimal coupling to gravity}.--- So far we have implicitly
assumed that the scalar field is minimally coupled to the gravity.
But the assumption of $\xi=0$, 
where $\xi$ is a nonminimal coupling to gravity ${\cal L}_\xi = -\xi \phi^2 {\cal R}/2$,
cannot be maintained in quantum field
theories since the parameter $\xi$ receives radiative
corrections \cite{Freedman:1974gs,Herranen:2014cua,Salvio:2014soa}.
The $\beta$ function of  $\xi$ is given by
$ \beta_\xi = ( \xi -1/6 ) \beta_{m^2}$,
where $\beta_{m^2}$ is the $\beta$ function of the mass term.
Hence $\xi$ gets renormalized unless $\xi=1/6.$
For example, in the minimal $B-L$ model \cite{Basso,IOO}, if we start from $\xi=0$ at the UV scale,
 a negative $\xi$ of order ${\cal O}(10^{-3})$ can be generated at the scale of inflation.
Then an effective mass term 
$m^2 \phi^2 /2$ with $m^2=12 \xi H^2$ is induced in the CW potential (\ref{CWpotential})
during  inflation with the Hubble constant $H$.

In the following, we study the small-field CW inflation with a negative mass term $m^2=12 \xi H^2<0.$ 
Since $\phi \ll M_{Pl}$,
the mass term is negligible compared to the original vacuum energy,
$m^2 \phi^2 \ll V_0.$
Hence, $V$ in the definitions of the slow roll parameters 
can be safely replaced by $V_0$. 
The first and the second derivatives of $V(\phi)$ are modified due to the mass term.
Since the inequality $\epsilon \ll |\eta|$ still holds, 
the condition $|\eta_{end}|=1$ determines the end of the inflation.
The slow roll parameter $\eta$ becomes
\be
\eta = 24 \left( \frac{M_{\rm Pl}}{M} \right)^2
\left( \frac{\phi}{M} \right)^2  \ln \frac{\phi^2}{M^2}
 + \frac{m^2 M_{Pl}^2}{V_0} \ . \label{nonminimal-eta}
\ee
The second term is constant; $m^2 M_{Pl}^2/V_0 = m^2/3 H^2$. 
It must be smaller than 1,
because otherwise the slow roll condition $|\eta|<1$ is always violated.
The inflation ends when the first term in $\eta$ grows over 1. Thus, $\phi_{end}$ is the same as in the CW inflation with no mass term (\ref{fieldvalue}).

The e-folding number is given by
\be
N&=&\frac{1}{M_{Pl}^2}\int^{\phi_{end}}_{\phi} \frac{V_0 d\phi}{A \phi^3 \ln(M^2/\phi^2) - m^2 \phi} 
\label{e-folding1}
\\
&\approx& 
\frac{1}{8|\xi|}
\left[ 
\ln \left( 
\frac{\tilde{A} \phi^2}{\tilde{A} \phi^2-m^2}
\right) \right]_\phi^{\phi_{end}}  \  .
\label{e-folding-xi}
\ee
where $\tilde{A}=A \ln(M^2/\phi_{c}^2).$  
In the second equality we performed the integration
by using an approximation that $\ln(M^2/\phi^2)$ is almost constant during the inflation.
The approximation is shown to be very good by comparing it with numerical calculations.

\begin{figure}[t]
 \begin{center}
   \includegraphics[width=0.9 \linewidth]{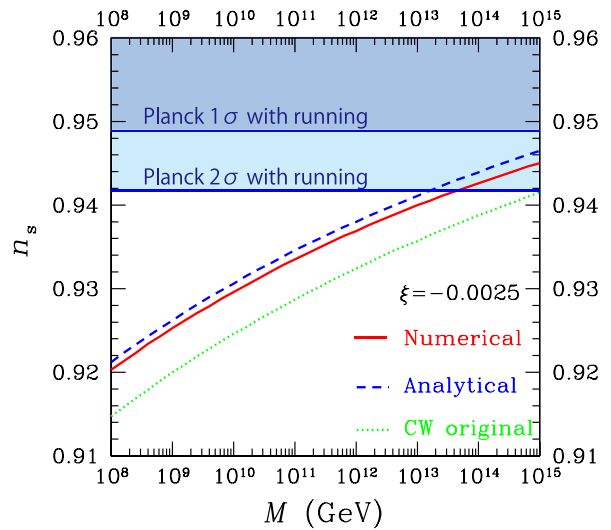}
   \caption{ Plots of the spectral index $n_s$ in the small-field CW inflation with the nonminimal coupling to gravity with $\xi = - 0.0025$.  The blue dashed line represents the analytical result
     (\ref{analytical-eta}) using an approximation.  The red solid
     line is based on a numerical calculation of the integral
     (\ref{e-folding1}).  The original CW result is plotted as the
     green dotted line for comparison. }
   \label{fig:ns-xi}
 \end{center}
\end{figure}

Combining  (\ref{nonminimal-eta}) and (\ref{e-folding-xi}), we
can express $\eta$ in terms of $N$ as
\be
\eta(N) \approx \frac{-12 \xi}{1-e^{-8 \xi N +12\xi}}+4 \xi \  . \label{analytical-eta}  
\ee
Since $\epsilon \ll |\eta|$, the spectral index is given by $n_s =1+ 2 \eta_{\rm CMB}.$
By using Eq.~(\ref{CMBefolding}), we can
rewrite the e-folding number $N$ in terms of the symmetry-breaking scale 
$M$. Here we assume $T_R=V_0^{1/4}$ for simplicity.
In Fig. \ref{fig:ns-xi},
we plot $n_s$  as a function of  $M$.
The dashed blue line is the analytical result (\ref{analytical-eta}) based on the approximation that the 
logarithmic term is constant.
The red solid line is the numerical result without using the above approximation.
We also plot the CW result without the nonminimal coupling $\xi=0$ (green dotted line) for comparison.
At fixed $M$ (equivalently at fixed $N_{\rm CMB}$), $n_s$  increases due to the effect of 
the non-minimal coupling to gravity by an amount of  $\sim 0.005$, 
but still $n_s<0.94$ in most of the region, $M<10^{13}$ GeV.
Therefore, the non-minimal coupling to gravity is 
 not sufficient to solve the small $n_s$ problem of the small-field CW inflation.
\\
\par
{\it Fermion condensates}.--- Another possibility to increase $n_s$ is
a generation of a linear term in the inflaton potential $V(\phi)$ due
to condensation of fermions coupled to the inflaton field.
We will
give two examples that may realize such a possibility (see also
Ref. \cite{Takahashi:2013cxa} for the origin of the linear term in
supersymmetric models).

First, in the $B-L$ model, 
the RH neutrinos $N_i$  are coupled to $\phi$ by $\phi \mbox{Tr} Y_N \bar{N}^c N.$ 
By integrating out $\phi$, four-Fermi interaction $G(\bar{N}^c N)(\bar{N}^c N)$ is induced with
$G \sim  Y_N^2/m_{\phi}^2$. If the Majorana Yukawa coupling is large enough, $Y_N \sim {\cal O}(10)$, 
the RH neutrinos may condense
 \cite{Barenboim:2010nm}.
Then the inflaton potential $V(\phi)$ acquires a linear term $-C \phi$, where $C=Y_N \langle \bar{N}^c N \rangle.$
The minus sign is a  convention to determine the direction of the linear potential.
A similar mechanism will work in a general model where the inflaton is coupled with strongly interacting fermions.
The effective potential of $\phi$ should be calculated with quantum corrections coming from effective propagating degrees of freedom which can differ from strongly interacting fermions themselves.

Another example is to use conventional chiral condensates of quarks.
When 
 the SM singlet scalar field $\varphi$ is mixed with the Higgs $h$ with a very small (and negative) mixing term 
$\lambda_{\rm mix} \varphi^2 h^2$, the potential $V(\varphi, h)$ has a valley along the direction
of $h^2=(|\lambda_{mix}|/2 \lambda_{h}) \varphi^2$ (see e.g. Ref. \cite{IOO}).
If the chiral condensate $\langle \bar{q} q \rangle \neq 0$ occurs near the origin of the potential,
it generates a linear term  $-C_0 h$ in the Higgs 
potential with $C_0 \sim   y  \langle \bar{q} q \rangle$, where $y$ is the Yukawa coupling.
Then the scalar mixing induces
a linear term \footnote{The Higgs field acquires a small VEV; 
$\langle h \rangle = (C_0/\lambda_h)^{1/3} \ll 246 {\rm GeV}. $ It breaks the EW symmetry near the origin 
of the potential 
and modifies the orbit of the classical motion on the $(h,\varphi)$ plane from the path along the valley.
If $C_0^{1/3} \gg \phi_{end}$, it may invalidate the generation of a linear term 
in the inflaton potential. Detailed studies  are left
for future investigations.
}
 in the direction of the valley $\phi$ with a coefficient
$C=\sqrt{|\lambda_{mix}|/2\lambda_h} C_0 =(246  /M[\mbox{GeV}]) C_0$. 
\\
\par
{\it CW inflation with fermion condensate}.---
In the following,  we suppose that  an appropriate
magnitude of a linear term exists in the inflaton potential (see Ref. \cite{Witten} for
a role played by the linear term in a different context). 
Then a constant term is added to the first derivative of $V$;
$
V'=A \phi^3 \ln (\phi^2/M^2)-C.
$
Since $\phi \ll M$,  the inequality $\epsilon \ll \eta$ still holds and
the inflation ends at the same value of the field $\phi_{end}$ as in the original CW inflation (\ref{fieldvalue}).
Since $V''$ is unchanged,
the field value at a fixed $\eta$ is independent of the value of $C$. 
But the relation between the e-folding number $N$  and the field value 
$\phi$ is modified as
\be
N=\frac{1}{M_{Pl}^2}\int^{\phi_{end}}_{\phi} \frac{V_0 d\phi}{A \phi^3 \ln(M^2/\phi^2) + C} \  .
\ee
The second term in the denominator reduces $N$ for fixed $\phi$.
It corresponds to the fact that, in the presence 
of the linear term,
$\phi_{\rm CMB}$ becomes smaller than in the original CW inflation, so that $|\eta|$ becomes smaller.
We parameterize $C$ as $C\equiv A \tilde{C} M^3
(M/M_{\rm Pl})^3$ for convenience.  Two terms in the denominator of
$N$ balance for $\widetilde{C} \sim 10^{-6} $ and $\phi_{\rm CMB}$,
or for $\widetilde{C} \sim 10^{-3}$  and $\phi_{end}$.  

\begin{figure}[t]
 \begin{center}
   \includegraphics[width=1.0 \linewidth]{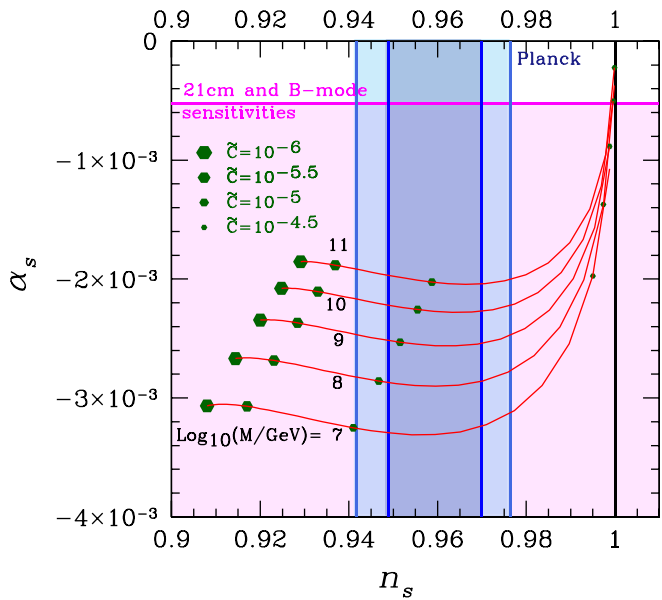}
   \caption{The relations between the spectral index $n_s$ and its running $\alpha$ 
   in the SFI model with a linear potential $C \phi$ are plotted.  Each red curve
     corresponds to a different symmetry-breaking scale $M = 10^{7-11}
     \ \mbox{GeV}$.  The relation of $N_{\rm CMB}$ and $M$ is
     given by Eq.~(\ref{CMBefolding}) with $T_R\sim V_0^{1/4}$.  A
     different $\tilde{C}$ corresponds to a different point on the red
     curve; $C\equiv A \tilde{C} M^3 (M/M_{\rm Pl})^3$. By the future 21 cm
     and CMB B-mode observations, the sensitivity on $\alpha_s$ will become $\delta \alpha_s = 5.3 \times
     10^{-4}$ \cite{Kohri:2013mxa}, which is denoted by the horizontal solid
     line. }
   \label{fig:ns}
 \end{center}
\end{figure}

In Fig. \ref{fig:ns}, we
plot the spectral index $n_s$ and its running $\alpha_s \approx - 2
\xi^{(2)}$ with $\xi^{(2)} \equiv V'V'''M^4_{\rm Pl}/V^2$  by
changing $\widetilde{C}$ for various values of $M$.  The scale of $M$
is varied between $10^7$ and $10^{11}$ GeV.  The predicted spectral index
$n_s$ becomes consistent with the observation when
$\widetilde{C} \sim 10^{-5}.$

To summarize, the linear term induced by the fermion condensate can solve the
small-$n_s$ problem in the small-field CW inflation.
The slow roll parameter $\epsilon$ is made bigger about $ 10$ times, but the inequality
$\epsilon \ll |\eta|$ still holds.  Thus the predicted
tensor-to-scalar ratio is negligibly small.  The magnitude of the
scalar perturbation again requires a very small quartic coupling
$A \sim 1.3 \times 10^{-14} (\tilde{C}/10^{-5})^2.$ In addition, as shown in Fig. \ref{fig:ns},
this model can be tested by the
future 21 cm and CMB B-mode observations on the running of the spectral
index \cite{Kohri:2013mxa}.
The preferred value of 
 $\tilde{C} \sim 10^{-5}$ corresponds to 
  the coefficient of the linear term along the valley
  $C \sim 10^{-19}M^3 (M/M_{\rm Pl})^3$.
In the model \cite{IOO} where the EW symmetry breaking is triggered by the $B-L$ gauge symmetry
breaking, $C$ is related to the chiral condensate of quarks 
$C_0 =y \langle \bar{q} q \rangle$ as 
$C=(246/M[{\rm GeV}]) C_{0}$. 
Hence, the condensation of the $B-L$ scalar $M$ and the chiral condensate $C_0$
are related as $M=(246 \times 10^{19} M_{\rm Pl}^3 C_0)^{1/7}$ in units of GeV.
Hence, the preferred value of the linear term in the inflaton potential
can be generated if we take, e.g.,  
$(C_0=y_u \langle \bar{u} u \rangle, M) = (10^{-5} \times (100 \ {\rm MeV})^3, 5 \times 10^{9}  \ {\rm GeV})$.\footnote{Near the origin of the potential, 
the Higgs field takes a small value, and the top quark is lighter than the 
QCD scale. Hence, if we instead use the
top quark condensation, the pair becomes
$(C_0=y_t \langle \bar{t} t \rangle, M) = ((100 \times {\rm MeV})^3, 2 \cdot 10^{10}  {\rm GeV}$).
}

{\it Reheating}.--- Finally, we evaluate the reheating temperature after the small-field CW inflation.
Here, we consider a particular model \cite{IOO}  mentioned in the above discussion of the fermion condensate 
where the SM singlet scalar field $\varphi$ with the CW potential and the SM Higgs boson $h$ are mixed by the interaction term $\lambda_{\rm mix} \varphi^2 h^2$ with a negative coupling constant.
A linear combination of these plays the role of an inflaton field $\phi$.
The Higgs mass is given as $m_h =\sqrt{|\lambda_{\rm mix}|} M$.
With a  small coupling $\lambda_{\rm mix}$, the damping rate of the oscillation of the inflaton $\phi$ is 
well approximated by the decay rate of $\varphi$ into the Higgs bosons:
\begin{align}
\Gamma_{\varphi \to h h}&\sim 10^{-2} \frac{(\lambda_{mix} M)^2 }{m_{\varphi}} = 10^{-2} \frac{m_{h}^4 }{m_{\varphi}M^2} \n
&\sim 10^{-2} \frac{m_{h}^3 }{M^2} \sim \left( \frac{10^2 {\rm GeV}}{M} \right)^2 {\rm GeV} \ .
\end{align}
In the second line, it is assumed that $m_\varphi \sim m_h$.\footnote{
This assumption is plausible in the classically conformal model with gauged $B-L$ symmetry:
Without the $B-L$ gauge coupling, $m_{\varphi}=\sqrt{\beta_{\varphi}} M$ must be smaller than $m_h$.
In the presence of the $B-L$ gauge coupling, $m_{\varphi}$ can be made larger, 
but at largest  $\sim m_h$. Otherwise, the Higgs mass tend to receive large quantum corrections from the mixing.
}
Comparing the decay rate to the Hubble expansion rate $H(T)/{\rm GeV}\approx 10^{-18}(T/{\rm GeV})^2 $,
we get the (maximum) reheating temperature
\begin{align}
T_{\rm rh}\sim \left( \frac{10^{11} {\rm GeV}}{M} \right) {\rm GeV} .
\end{align}
For successful baryogenesis, the reheating temperature must be higher than $T_{\rm EW}\sim 10^2 {\rm GeV}$, and thus the minimum of the CW potential  must be located at $\varphi = M<10^9 {\rm GeV}$.\footnote{
In the gauged $B-L$ model, the Majorana masses of the right-handed neutrinos $N_{i}$ are generated through the Yukawa coupling $Y_N \varphi N \overline{N}$, and  $\varphi$ can decay into $N \overline{N}$.
However, it turns out that the decay rate $\Gamma_{\varphi \to N \overline{N}}\sim 10^{-2} Y_N^2 m_{\phi}\sim 10^{-2} m_N^2 m_{\phi} /M^2 < 10^{-2} m_{\phi}^3/M^2$ is smaller than $\Gamma_{\varphi \to hh} \sim 10^{-2} m_{h}^3/M^2$.
Taking into account an effect of the decay process of $N$ to the SM particles, 
the $\varphi \to N \overline{N}$ channel cannot increase the reheating temperature evaluated above.
Of course, the decay channel becomes important when $\varphi$ cannot decay into the Higgs bosons.
}

{\it Summary}.--- In this paper, we studied effects of a nonminimal
coupling to gravity and fermion condensation on the small-field CW
inflation.  The original small-field CW inflation predicts a rather
small spectral index $n_s <0.94$, compared to the Planck measurement
$n_s=0.942-0.978$ \cite{Ade:2013zuv} in the case with the running of $n_s$.  The effect of a nonminimal coupling to gravity with a
negative value $|\xi| \sim {\cal O}(10^{-3})$ is shown to increase the
spectral index $n_s$ by 0.005 but is not sufficient to reconcile the
prediction with the data for $M<10^{13} {\rm GeV}$.
We then studied the effect of the condensation
of fermions coupled to the inflaton field $y _\psi \phi \bar\psi
\psi.$ If $\langle \bar\psi \psi \rangle \neq 0,$ a linear term is
generated in the inflaton potential $V(\phi).$ In particular, when the
inflaton and the Higgs are mixed, the chiral condensate of quarks
induces a linear term in the inflaton potential.  We showed that an
appropriate magnitude of condensation can make the theoretical
prediction of $n_s$ consistent with the observational data.  The
tensor-to-scalar ratio $r=16 \epsilon$ is negligibly small. However,
the running of the spectral index will be tested by
the the future 21 cm and CMB B-mode observations \cite{Kohri:2013mxa}.
\\
\begin{acknowledgments}
{\it Acknowledgments}.--- The authors would like to thank M. Kawasaki, R. Kitano, T. Moroi, K. Nakayama and N. Yamada
for valuable comments and discussions.  
This work is supported by the Grant-in-Aid for Scientific Research from the
Ministry of Education, Science, Sports, and Culture, Japan,
No. 23540329, No. 23244057 (S.I.), No. 23540327, No. 26105520, and 26247042
(K.K.).  This work is also supported by ``The Center for the
Promotion of Integrated Sciences (CPIS)'' of Sokendai (1HB5804100).
\end{acknowledgments}

\end{document}